\documentclass[12pt,aps,prd,preprint,tightenlines,amsmath,amssymb,showpacs,nofootinbib]{revtex4}
\pdfoutput=1

\usepackage{amsmath}         
\usepackage{amssymb}         
\usepackage{amsfonts}        
\usepackage{graphicx}        
\usepackage{wasysym}
\usepackage{tikz}
\usepackage[compat=1.1.0]{tikz-feynman}
\usepackage[utf8]{inputenc}

\usepackage{color}         
\usepackage{slashed}       
\usepackage{enumerate}		%

\graphicspath{{figures/}}	

\renewcommand{\vec}[1]{\mathbf{#1}} 

\let\oldenumerate\enumerate
\renewcommand{\enumerate}{
  \oldenumerate
  \setlength{\itemsep}{1pt}
  \setlength{\parskip}{0pt}
  \setlength{\parsep}{0pt}
}

\let\olditemize\itemize
\renewcommand{\itemize}{
  \olditemize
  \setlength{\itemsep}{1pt}
  \setlength{\parskip}{0pt}
  \setlength{\parsep}{0pt}
}

\newcommand{\PRE}[1]{{#1}}  

\newcommand{\mev}{\text{MeV}}
\newcommand{\gev}{\text{GeV}}
\newcommand{\tev}{\text{TeV}}

\renewcommand{\eqref}[1]{Eq.~(\ref{#1})}

\newcommand{\figref}[1]{Fig.~\ref{fig:#1}}

\newcommand{\lag}{\mathcal L}

\def\be{\begin{equation}}
\def\ee{\end{equation}}
\def\bea{\begin{eqnarray}}
\def\eea{\end{eqnarray}}

\usepackage[colorlinks=true,citecolor=blue,linkcolor=blue,
urlcolor=blue,hypertexnames=false]{hyperref}

\usepackage{microtype}				

\begin{document}

\preprint{UCI-TR-2018-16}

\title{
{
Self-Interacting Dark Matter With a Neutrinophilic Scalar Mediator
}
}

\author{Arvind Rajaraman\footnote{{arajaram@uci.edu}}
}
\affiliation{Department of Physics and Astronomy, University of
	California, Irvine, California 92697, USA
	\PRE{\vspace*{.15in}}
}

\author{Jordan Smolinsky\footnote{{jsmolins@uci.edu}}
}
\affiliation{Department of Physics and Astronomy, University of
  California, Irvine, California 92697, USA 
\PRE{\vspace*{.15in}}
}


\begin{abstract}
\PRE{\vspace*{.2in}}
\noindent

We  examine the phenomenology of  a simplified model of 
fermionic dark matter  
coupled to a light scalar mediator carrying lepton number 2. 
We find that the mediator can be very light and still consistent with 
laboratory and cosmological bounds. This model satisfies the thermal relic condition for 
natural values of dimensionless coupling constants and admits a 
mediator in the $10 - 100 ~\text{MeV}$ mass range favored by small 
scale structure observations. As such, this model provides an
excellent candidate for self-interacting dark matter. 
\end{abstract}


%
%
\maketitle

\section{Introduction}

The nature of dark matter remains among the most prominent 
questions in fundamental physics, and one of the best motivations for models of 
physics beyond the Standard Model (SM). While dark matter (DM) was discovered by its 
gravitational effects \cite{Rubin:1970zza,Rubin:1980zd,Clowe:2006eq}, it is 
still empirically unknown whether it participates in any other fundamental interaction. 

The best-motivated theories of physics beyond the Standard Model, supersymmetric 
extensions of the SM, yield weakly interacting cold dark matter candidates at the weak scale 
\cite{Jungman:1995df}, and a great deal of effort has been focused on
searching for such dark matter candidates. However, many recent
astrophysical observations have cast doubt on these models, since these models 
appear to be in tension with various observations of the inner halos of galaxies. 
This has led to the suggestion that dark matter
in fact has large self interactions, and self-interacting
dark matter can indeed solve many of these problems.
\cite{Bernal:2015ova,Balducci:2017vwg,Ren:2018jpt}.
In light of these small scale 
structure observations, it is of great interest  to consider
models of dark matter coupled to a light boson, like a 
dark photon \cite{Holdom:1985ag,Morrissey:2009ur} or dark higgs 
\cite{Patt:2006fw,MarchRussell:2008yu} 
which can produce a large dark matter  scattering cross section.

In this work, we will examine the phenomenology of a simplified model of 
fermionic dark matter 
coupled to a light complex scalar $\phi$ carrying lepton number 2. 
Such a light particle might be visible through its interactions with the Standard Model.
Dark sector particles may be produced in colliders 
and found through their missing energy signatures 
\cite{ATLAS:2012ky,Aad:2012awa,Aad:2012fw,Chatrchyan:2012me,Chatrchyan:2012tea,Goodman:2010ku, Zhou:2013fla}, 
they may scatter off of SM detector constituents to produce an observable 
recoil \cite{Aprile:2018dbl}, or they may annihilate or decay to produce a 
flux of energetic SM particles \cite{Aartsen:2017ulx}. Finally, a model of dark matter 
 must satisfy the combined constraints of all applicable 
laboratory tests and predict a cosmological abundance consistent with 
observations \cite{Aghanim:2018eyx}.

As we show in the next section, the neutrinophilic scalar portal model is a completely viable model of 
dark matter.
The appropriate relic density is obtained through the coupling of the dark matter
to the neutrinos. Other constraints are weak;
indeed, such a model is hard to constrain, since
even a very light scalar coupled only to the neutrinos and dark matter
has relatively few signals \cite{Campo:2017nwh,Primulando:2017kxf}

We also analyze the case when there are further 
interactions 
between the light scalar and the quarks of the Standard Model. 
The symmetries force such couplings to be 
nonrenormalizable.  The interactions 
facilitated by these nonrenormalizable operators can be probed by colliders and 
direct detection experiments.
We show that
the current bounds on these interactions are very weak, even if the mediator
is very light. This then shows that the neutrinophilic scalar portal can naturally accommodate 
self interacting
dark matter, 
over a wide range of dark matter and mediator masses.

\section{A Simplified Model of a Neutrinophilic Scalar Mediator}
We consider a model of dark matter, where the 
dark matter is a  Majorana fermion $\chi$ of mass $m_\chi$.
It is coupled to a scalar (the neutrinophilic scalar) of mass $m_\phi$.
The neutrinophilic scalar carries lepton number, so that its only tree-level interactions with the 
Standard Model come through coupling to the neutrino majorana mass. We suppose that the 
dark matter interaction with the neutrinophilic scalar follows the same structure 
(i.e. the dark matter effectively has lepton number),
 so that the leading 
interactions of the theory may be written
\begin{align}
\lag_\text{ren} = - g_\nu \overline{\nu_L^c} \nu_L \phi - 
g_\chi \overline{\chi^c} \chi \phi + \text{h.c.} \ , 
\end{align}
where $g_i$ are dimensionless coupling constants.

We may fix the couplings through the thermal relic condition, following 
the procedure and notation of \cite{Gondolo:1990dk}. 
There are two annihilation 
channels we need to consider.

\begin{figure}[h] 
	\hspace*{-.5cm}
	\includegraphics[width=0.45\linewidth]{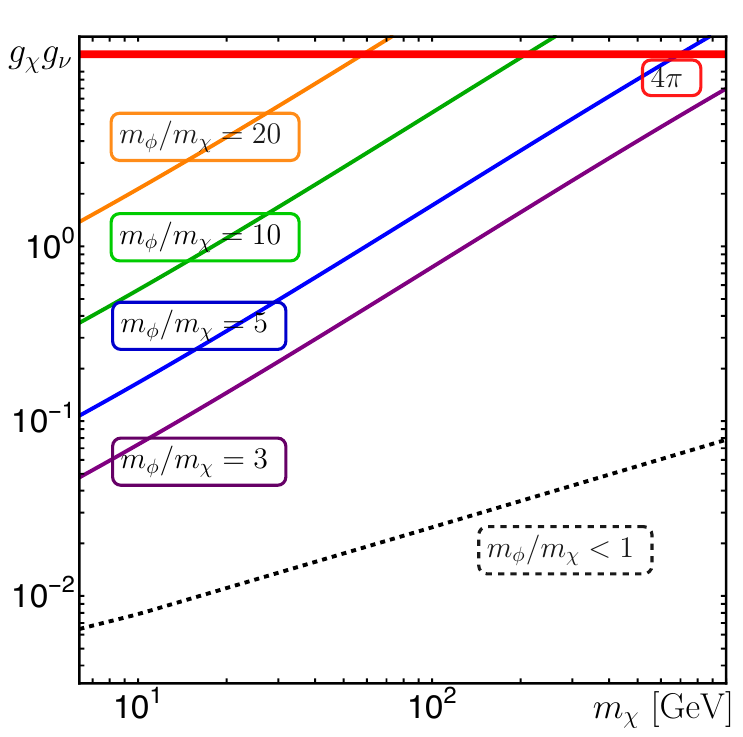}
	\vspace*{-0.1in} 
	\caption{Thermal relic constraint on dark matter and neutrino coupling of the neutrinophilic scalar as a 
	function of dark matter mass $m_\chi$, for indicated values of $m_\phi/m_\chi > 1$. \textbf{Dashed:} Relic 
	constraint for $m_\phi/m_\chi < 1$, $g_\nu = 0.1$, only weakly sensitive to changes in $m_\phi$.
	\textbf{Red:} Perturbative upper limit on $g_\chi g_\nu$.}
	\label{fig:thermrelic}
	\vspace*{-0.1in}
\end{figure}

If  $m_\phi > m_\chi$, 
the dominant process will be $\chi \chi \rightarrow \nu \nu$, which is p-wave:

\begin{center}
	\begin{tikzpicture}
	\begin{feynman}
	\vertex (a1) {\(\chi\)};
	\vertex[right=1.5cm of a1] (a2);
	\vertex[right=1.5cm of a2] (a3) {\(\nu\)};
	
	\vertex[below=3em of a1] (b1) {\(\chi\)};
	\vertex[right=1.5cm of b1] (b2);
	\vertex[right=1.5cm of b2] (b3) {\(\nu\)};
	
	\vertex[below=1.5em of a2] (c2) ;
	\vertex[right=0.5cm of c2] (c3);
	\vertex[left=0.5cm of c2] (c1);

	\diagram* {
		{[edges=fermion]
			(a1) -- (c1)  -- (b1),
			(a3) -- (c3) -- (b3)
		},
		(c1) -- [charged scalar, edge label=\(\phi\)] (c3)
		
	};

	\end{feynman}
	\end{tikzpicture}
\end{center}
\begin{equation}
\langle \sigma_{\chi \chi \rightarrow \nu \nu} v \rangle = \frac{3 g_\chi^2 
g_\nu^2}{4 \pi m_\chi^2} \frac{ (1 - m_\nu^2/m_\chi^2)^{3/2}}{ (4- m_\phi^2/m_\chi^2)^2}  
\frac{1}{x} + \mathcal{O}(x^{-2}) \ ,
\end{equation}
where $x = m_\chi/T$, with $T$ the temperature. In this regime the thermal relic values 
of $g_\chi g_\nu$ will be determined both by $m_\chi$ and the ratio $m_\chi/m_\phi$. This 
dependence is shown in \figref{thermrelic}. In this figure we have omitted analysis of 
the resonant regime $m_\phi - 2 m_\chi \ll m_\chi$, in which the thermal relic target 
may be depressed by several orders of magnitude.

 Secondly, the process $\chi \chi \rightarrow \phi \phi$ 
dominates the dark matter annihilation in the regime $m_\chi > m_\phi$. Its thermal 
averaged cross section is s-wave:
\begin{center}
	\begin{tikzpicture}
	\begin{feynman}
	\vertex (a1) {\(\chi\)};
	\vertex[right=1.5cm of a1] (a2);
	\vertex[right=1.5cm of a2] (a3) {\(\phi\)};
	
	\vertex[below=3em of a1] (b1) {\(\chi\)};
	\vertex[right=1.5cm of b1] (b2);
	\vertex[right=1.5cm of b2] (b3) {\(\phi\)};
		
	\diagram* {
		{[edges=fermion]
			(a1) -- (a2) -- (b2) -- (b1)
		},
		(a2) -- [charged scalar] (a3),
		(b3) -- [charged scalar] (b2),
		
	};
	
	\end{feynman}
	\end{tikzpicture}
\end{center}
\begin{equation}
\langle \sigma_{\chi \chi \rightarrow \phi \phi} v \rangle = \frac{g_\chi^4}{16 \pi m_\chi^2 } \frac{(1-m_\phi^2/m_\chi^2)^{3/2}}{(2-m_\phi^2/m_\chi^2)^2} + \mathcal{O}(x^{-1}) \ .
\end{equation} 
Notice that the annihilation $\chi \chi \rightarrow \phi \phi$ has only weak 
dependence on $m_\phi$, so that in the regime that this channel dominates, $g_\chi$ may 
be determined completely by the mass $m_\chi$ of the dark matter itself. Explicitly, the 
thermal relic condition in this regime furnishes the relation 
\begin{equation}
\alpha_\chi \equiv g_\chi^2/4\pi \approx 0.07 ~m_\chi/\tev ~~~(m_\chi > m_\phi)\ . 
\end{equation}
Note that the mediator can be very light in this scenario. Note also that $g_\nu$ is not directly constrained by the thermal relic condition in the regime $m_\phi < m_\chi$.

We therefore see that the thermal relic density condition can be satisfied
over an enormous rage of parameter space, including relatively weak 
masses for the neutrinophilic scalar mediator. 

Furthermore, our preceding calculations
have assumed the standard cosmology. Nonstandard thermal histories for the universe
may significantly change our result. For instance
quintessence models generating an early phase of kination dominance, may
satisfy the thermal relic constraint even with an annihilation cross section
up to three orders of magnitude larger \cite{Pallis:2005hm}. In such cosmologies
 $g_\chi$, $g_\nu$ could be much larger. We leave an analysis of this possibility
 to future work.

There are few other constraints on this model. The main one comes from indirect detection.
However, because the primary final state of neutrinophilic scalar portal dark matter annihilation is 
neutrinos, it 
is difficult  to set meaningful constraints on the dark matter 
annihilation. IceCube furnishes the strongest indirect detection limits on dark 
matter annihilation to neutrinos, but these do not exclude the thermal relic 
cross section \cite{Aartsen:2017ulx}.

\section{Nonrenormalizable interactions}
To further probe experimental constraints on this model, we must
enlarge our model to include couplings of the mediator to quarks and charged leptons.

There are no renormalizable couplings allowed between the mediator and any charged SM 
fermions, so we must introduce nonrenormalizable couplings.
We will 
restrict our attention to dark matter coupling through either 
a scalar or a pseudoscalar quark current, and following the 
principle of minimal flavor violation the coupling constants to these currents will 
be taken to be proportional to the quark masses. 
We therefore have 
\bea
\lag_\text{}&=&\lag_\text{ren}+\lag_\text{nonren}
\eea
where
\bea
\lag_\text{ren}&=&- g_\nu \overline{\nu_L^c} \nu_L \phi - 
g_\chi \overline{\chi^c} \chi \phi + \text{h.c.} \ , 
\eea
For the scalar current coupling, we take
\bea
\lag_\text{nonren} &=& \frac{1}{M_*^2} \sum_q m_q \phi^* \phi \overline{q}q
\eea
and for the pseudoscalar current coupling, we take
\bea
\lag_\text{nonren} &=& 
\frac{1}{M_*^2} \sum_q i m_q \phi^* \phi \overline{q} \gamma^5 q
\eea
Note that the nonrenormalizable interactions correspond to C1 and C2
in the naming convention of \cite{Goodman:2010ku}.

Here $M_*$ is  a scale associated with the UV completion of this theory. 
While the non-discovery of new physics at the LHC might suggest
that this new physics should be at least at a TeV, we shall remain agnostic, and 
not impose any theoretical prejudice on the parameters.
 The parameter 
space of this theory is then spanned by the parameters 
$m_\chi$, $m_\phi$, $g_\nu$, $g_\chi$, and $M_{*}$. 
 We now map out the constraints that may be placed on this parameter space by 
colliders and direct detection.

\subsection{Collider Constraints}

Because of the structure of the scalar interaction with the quarks, production of the 
dark matter at colliders is suppressed by a loop and a factor of $g_\chi^2$ at the 
amplitude level. The dominant process observable at the LHC experiments ATLAS and 
CMS is then $ p p \rightarrow \phi \phi^* + X$, where $X$ is any SM final state. These 
events are marked by $X$ recoiling against the invisible pair of mediators, which do not 
interact with particle detectors at the interaction point. 

\begin{figure}[t] 
	\hspace*{-.5cm}
	\includegraphics[width=0.45\linewidth]{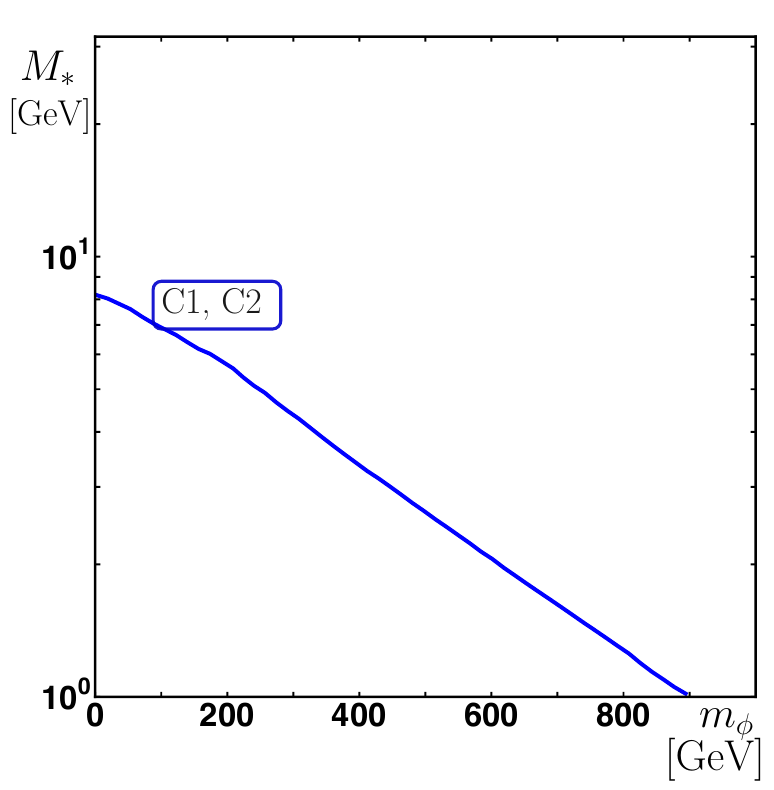}
	\vspace*{-0.1in} 
	\caption{Combined limit on $\Lambda$ from ATLAS monojet \cite{ATLAS:2012ky}, 
	mono-$\gamma$ \cite{Aad:2012fw}, mono-Z \cite{Aad:2012awa,Carpenter:2012rg}, and 
	CMS monojet \cite{Chatrchyan:2012me} and mono-$\gamma$ \cite{Chatrchyan:2012tea} 
	searches with $\sqrt{s}= 7~\tev$ \cite{Zhou:2013fla}}.
	\label{fig:C1limit}
	\vspace*{-0.1in}
\end{figure}

Leading limits on $\phi \phi^*$ production come from consideration of monojet + 
$\slashed{E_T}$ events at ATLAS and CMS. The largest background contribution is from
 a jet recoiling against an off-shell Z boson that decays to neutrinos. In order to reduce 
this background the ATLAS search considers lepton-less events with a missing transverse 
energy of $\slashed{E_T} > 350 ~\gev$ and a primary jet $p_T > 350 ~\gev$. The 
companion CMS analysis allows a lower primary jet $p_T > 110 ~\gev$ while placing 
the same $\slashed{E_T}$ cut. Combined limits from these monojet searches 
\cite{ATLAS:2012ky,Chatrchyan:2012me}, as well as mono-$\gamma$ 
\cite{Aad:2012fw,Chatrchyan:2012tea}, and mono-$Z$ \cite{Aad:2012awa,Carpenter:2012rg} 
at the $\sqrt{s}=7~\tev$ LHC are presented in \cite{Zhou:2013fla}, we reproduce their 
results in \figref{C1limit}. 

We see that colliders place relatively weak limits on 
$M_*$, owing to the momentum-independent contact interaction between the scalar and 
the quarks and the lack of direct coupling to gluons.
These constraints will become stronger if and when the current and future
data from the LHC 
are used to constrain the monojet signature.  Independent of the couplings 
between the mediator and the dark matter or neutrinos, we see that the neutrinophilic scalar is 
currently a 
viable mediator at nearly all masses, as long as $M_* \gtrsim 10~\gev$.

\subsection{Direct Detection}

As the Earth traverses the dark matter halo, dark matter particles may scatter off of 
heavy nuclear targets, producing an observable recoil spectrum. In this model, the 
dominant contribution to direct detection occurs through the t-channel exchange of 
two scalars with the SM target. 
In the case that the mediator-quark interaction is 
described by operator C1, the dark matter scattering is spin-independent and the 
leading limits on its cross section come from XENON-1T \cite{Aprile:2018dbl}. Dark 
matter scattering through the C2 operator is spin-dependent and is most strongly 
constrained by LUX exclusions \cite{Akerib:2017kat}. 

In this section we present 
a calculation of the relevant cross sections for both types of mediator-quark interaction 
and the resulting limits on the model parameter space. Interactions between the dark 
matter and the SM fermions are automatically suppressed to one loop order, which weakens the
bounds.

\subsubsection{Scalar current}
The matrix element for the direct detection scattering process 
$\chi q \rightarrow \chi q$ with operator C1 is given by
\begin{center}
	
	\begin{tikzpicture}
	\begin{feynman}
	\vertex (a1) {\(\chi\)};
	\vertex[right=1cm of a1] (a2);
	\vertex[right=1cm of a2] (a3);
	\vertex[right=1cm of a3] (a4) {\(\chi\)};
	
	\vertex[below=3em of a1] (b1) {\(q\)};
	\vertex[right=1.5cm of b1] (b2);
	\vertex[right=1.5cm of b2] (b3) {\(q\)};

	\diagram* {
		{[edges=fermion]
			(a1) -- (a2) -- (a3) -- (a4),
			(b1) -- (b2) -- (b3) ,
		},
		(a2) -- [charged scalar, edge label'=\(\phi\)] (b2),
		(b2) -- [charged scalar, edge label'=\(\phi\)] (a3),

	};

	\end{feynman}
	\end{tikzpicture}
\end{center}

\begin{figure}[t] 
	\hspace*{-.5cm}
	\includegraphics[width=0.45\linewidth]{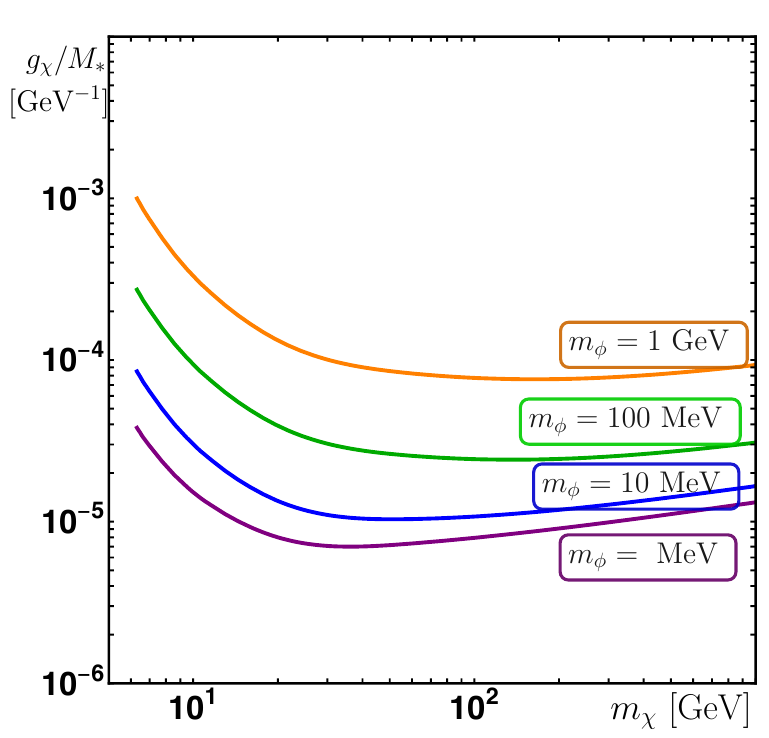} \qquad
	\includegraphics[width=0.45\linewidth]{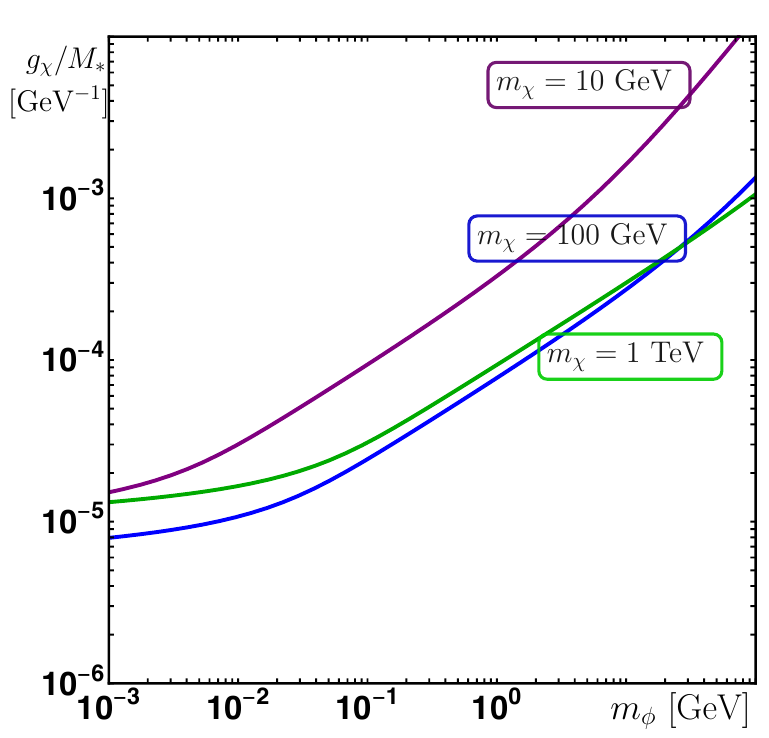} \\
	\vspace*{-0.1in} 
	\caption{\textbf{Left:} Direct detection limits on majoron couplings from 
	XENON-1T \cite{Aprile:2018dbl} as $m_\chi$ varies, with indicated majoron masses. 
	\textbf{Right:} Same, but for indicated values of $m_\chi$ as $m_\phi$ varies.}
	\label{fig:resultsSI}
	\vspace*{-0.1in}
\end{figure}

In order to build up the nuclear cross section from the partonic matrix element above, we 
follow the procedure of \cite{Backovic:2015cra}. The matrix element for direct detection 
factorizes into a universal, dark-matter related piece which we call $\alpha_q$, and a 
target-dependent Standard Model piece
\begin{equation}
\langle \mathcal{M} \rangle = \alpha_q \langle \bar{\psi}_q \psi_q \rangle \ ,
\end{equation}
where $\alpha_q$ is found after a loop calculation to be 
\begin{align}
\begin{split}
\alpha_q &= \frac{g_\chi^2 m_q [\bar{u}_3 u_1]}{M_*^2}\frac{m_\chi}{4m_\chi^2 - t} 
\big[2 B_0(p_1 - p_3, m_\phi, m_\phi) - B_0(p_1, m_\phi, m_\chi) \\ &~~~ - 
B_0(p_3, m_\chi, m_\phi) + (8 m_\chi^2 - 2 m_\phi^2 - t) C_0(p_1, -p_3, m_\phi, 
m_\chi, m_\phi)\big] \ .
\end{split}
\end{align}
Here $B_0, C_0$ are the Passarino-Veltman functions \cite{Ellis:2011cr}. Note that $\alpha_q$ is the same for 
both operators we will consider; 
the scalar/pseudoscalar nature of the mediator-quark operators will only affect 
scattering at the level of the nuclear form factors. 
The matrix element for the dark matter interacting with a nucleon through 
the scalar current is then
\begin{equation}
f_N^S = m_N \sum_{q = u, d, s} \frac{\alpha_q}{m_q} f_N^{Sq} + \frac{2}{27} 
m_N f_N^{SG} \sum_{q = c, b, t} \frac{\alpha_q}{m_q} \ ,
\end{equation}
where the numerical values of the form factors $f_N^{S q}$ are \cite{Alarcon:2011zs,Alarcon:2012nr}
\begin{align}
\begin{split}
f_p^{Su} &= 0.021,   f_n^{S u} = 0.019 \\
f_p^{Sd} &= 0.041,   f_n^{Sd} = 0.045 \\
f_p^{Ss} &= 0.017,   f_n^{Ss} = 0.017 \ ,
\end{split}
\end{align}
and $f_N^{SG} = 1 - \sum_q f_N^{Sq}$. Combining the interactions with individual nucleons 
into the nuclear cross section yields
\begin{equation}
\sigma_{\text{SI}} = \frac{4}{\pi} \mu_A^2 \left[ Z f_p + (A-Z) f_n \right]^2
\end{equation}
where $\mu_A$ is the reduced mass of the dark matter-nucleus system.
Limits on the spin-independent cross section of dark matter-Xenon scattering from 
XENON-1T \cite{Aprile:2018dbl} may now be directly translated into limits on 
$g_\chi/M_*$. These limits are shown in \figref{resultsSI}.

\subsubsection{Pseudoscalar Current}
Now we consider the case of the pseudoscalar operator. The quark-level operator C2 hadronizes to the 
pseudoscalar nuclear current
\begin{equation}
\frac{m_q}{M_*^2} \overline{q} i \gamma^5 q \rightarrow c_N \frac{m_N}{M_*^2} 
\overline{N} i \gamma^5 N \ ,
\end{equation}
with
\begin{equation}
c_N = \sum_{q = u, d, s}  \left[1 - 6 \frac{\overline{m}}{m_{q}} \right] 
\Delta_q^{(N)} \ ,
\end{equation}
where $\overline{m} \equiv \left[\sum_{q = u, d, s} m_q^{-1}\right]^{-1}$ and 
$\Delta_q^{(N)}$ are the quark spin contents of a nucleon, with numerical values 
\cite{Cheng:2012qr}:
\begin{align}
\begin{split}
\Delta_u^{(p)} &= \Delta_d^{(n)} = 0.84 \\
\Delta_d^{(p)} &= \Delta_u^{(n)} = -0.44 \\
\Delta_s^{(p,n)} &= -0.03 \ .
\end{split}
\end{align}

\begin{figure}[t] 
	\hspace*{-.5cm}
	\includegraphics[width=0.45\linewidth]{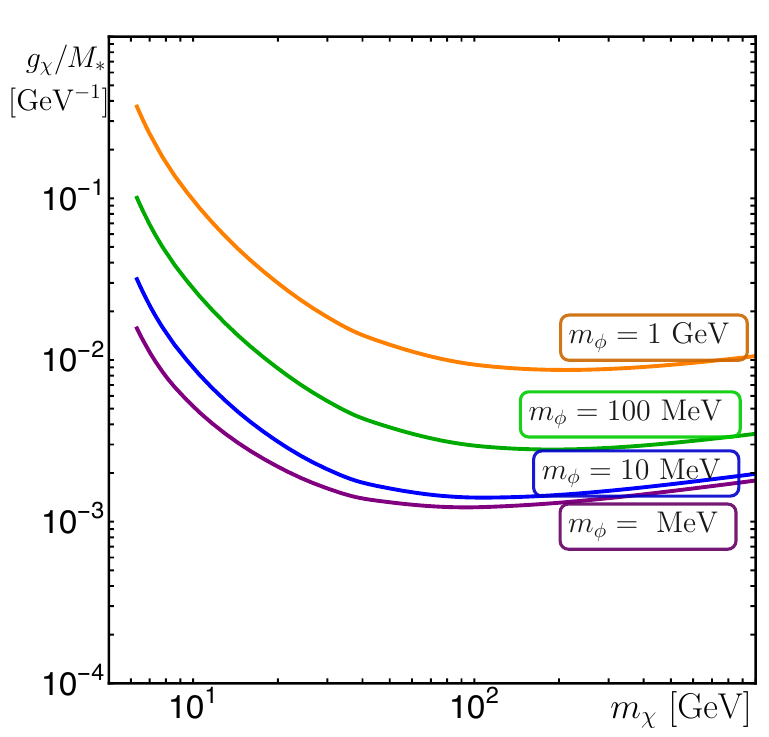} \qquad
	\includegraphics[width=0.45\linewidth]{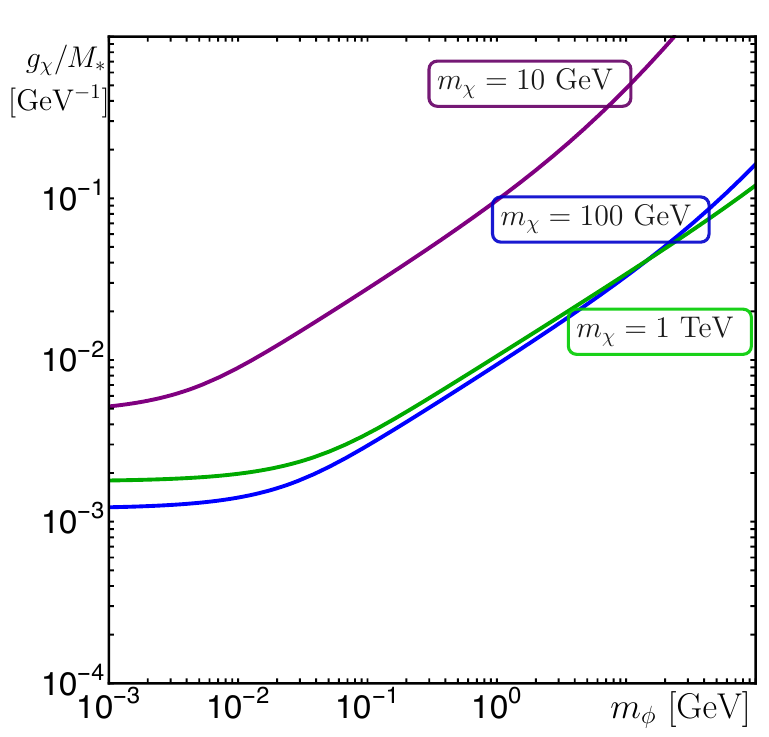} \\
	\vspace*{-0.1in} 
	\caption{\textbf{Left:} Spin-dependent direct detection limits on majoron-quark 
	couplings from LUX \cite{Akerib:2017kat} as $m_\chi$ varies, with indicated 
	majoron masses. \textbf{Right:} Same, but for indicated values of $m_\chi$ as 
	$m_\phi$ varies}
	\label{fig:resultsSD}
	\vspace*{-0.1in}
\end{figure}

In the nonrelativistic limit, the nuclear current reduces to \cite{Fitzpatrick:2012ix}
\begin{equation}
\overline{N} i \gamma^5 N \rightarrow -2 i \vec{S}_N \cdot \vec{q} \ ,
\end{equation}
and the corresponding nuclear spin-averaged transition probability is given in terms of 
the nuclear form factors
\begin{equation}
\frac{1}{2j+1}\sum_{\text{spins}} \vert \langle A \vert \sum_N  i c_N \vec{S}_N \cdot 
\vec{q} \vert A \rangle \vert^2 =  \frac{m_A^2}{m_N^2} \sum_{N,N'=p,n} 
c_N c_{N'} ~F_{10,10}^{(N,N')} (v^2, q^2) \ ,
\end{equation}
where $m_A$ is the mass of the target nucleus with mass number $A$ and 
$F^{(N,N')}_{10,10} = q^2 F_{\Sigma''}^{(N,N')}/4$ with $F_{\Sigma''}^{(N,N')}$ the 
axial longitudinal response function, tabulated for various nuclei in 
\cite{Fitzpatrick:2012ix}. We will consider the bounds on spin-dependent dark matter 
scattering from the LUX experiment, and as such use the form factors for $^{129}$Xe and 
$^{131}$Xe, weighted by their relative isotopic abundances, to calculate the predicted 
cross section for $N=p,n$. Spin-dependent direct detection limits on the proton and neutron 
cross section may be combined \cite{Tovey:2000mm} according to
\begin{equation}
\left(\sqrt{\frac{\sigma_p^\text{th}}{\sigma_p^\text{lim}}} + 
\sqrt{\frac{\sigma_n^\text{th}}{\sigma_n^\text{lim}}} \right)^2 > 1 \ ,
\end{equation}
where $\sigma_N^\text{lim}$ is the empirical upper limit on the WIMP-nucleon cross 
section and $\sigma_N^{\text{th}}$ is the model prediction. We use the spin-dependent 
cross section upper limits from LUX \cite{Akerib:2017kat} in order to bound the 
combination $g_\chi/M_*$ as shown in \figref{resultsSD}.

The main result from these analysis is that extremely small values of the mediator mass (as
small as an MeV) are viable.

\section{Conclusion}
In this paper we have constructed a model of neutrinophilic scalar mediators, and showed that 
in this model, the mediator field can be 
very light, allowing for the possibility of a 
self-interacting dark sector.
We found that the dark matter reproduces 
the observed relic density using thermal freeze out with natural values of 
dimensionless coupling constants.

We also analyzed the experimental constraints on interactions of the
neutrinophilic scalar with charged Standard Model 
fermions 
coming from collider and direct detection experiments. 
We found that extremely light mediator masses were viable;
the scalar could exist in the $10-100~\mev$ range favored by small scale structure 
observations without being excluded by colliders, and that weak scale dark matter coupling 
to this scalar is viable if the high scale new physics facilitating 
spin-independent (-dependent) direct detection occurs above 
$\sim 10~\tev$ ($\sim 100 ~\gev$). 

Neutrinophilic scalar mediators are therefore an excellent candidate for
a theory of  self-interacting dark matter.  At the same time,
ongoing experiments at the LHC and future experiments like BELLE 2 will
 will further constrain this 
model of dark matter, either discovering these scalars or ruling out 
larger regions of parameter space. 
It would be very interesting to
analyze the cosmology of these models and investigate whether the issues with small scale 
structure can be solved;
we will perform this analysis in future work.

\section*{Acknowledgments}

This work is supported by NSF Grant No.~PHY--1620638.  
%
%
Numerical calculations were
performed using \emph{Mathematica 11.1}~\cite{Mathematica10}. Feynman diagrams were drawn using TikZ-Feynman \cite{Ellis:2016jkw}.

\appendix*

\bibliography{bibmajoron}

\end{document}